Failure Analysis and Reliability Evaluation of Modulation Techniques for Neutral Point Clamped Inverters—A Usage Model Approach

Masoud Farhadi, Mehdi Abapour, Mehran Sabahi

# Failure Analysis and Reliability Evaluation of Modulation Techniques for Neutral Point Clamped Inverters—A Usage Model Approach

Masoud Farhadi [1], Mehdi Abapour [2*], Mehran Sabahi [3]

[1] Department of Electrical and Computer Engineering, University of Tabriz, 29 Bahman Blvd. Tabriz, Iran
[2] Department of Electrical and Computer Engineering, University of Tabriz, 29 Bahman Blvd. Tabriz, Iran
[3] Department of Electrical and Computer Engineering, University of Tabriz, 29 Bahman Blvd. Tabriz, Iran
[*] corresponding.abapour@tabrizu.ac.ir

**Abstract:** Up to now, many modulation techniques have been proposed for neutral point clamped (NPC) inverters. In this paper, for the first time, a general methodology is applied to calculate and compare the failure analysis and reliability of NPC inverter with most commonly used control strategies. Also, the mean time to failure (MTTF) of NPC inverter is derived for different control strategies. It is demonstrated that the key feature of control strategies in determining the reliability of inverter is their loss distribution among the switches. The failure rate of components that is relevant to this study and junction temperature calculation is developed, then conduction losses and switching losses of switches for different control strategies are calculated. Finally, the most reliable control strategy is identified. Experimental results obtained have promptly justified the theoretical analysis and outlined procedure.

**Nomenclature**

| | |
|---|---|
| $\lambda_{FET}$ | Failure rate of MOSFET (FIT = $10^{-9}$ failure/hour). |
| $\lambda_D$ | Failure rate of diode (FIT). |
| $\lambda_C$ | Failure rate of capacitor (FIT). |
| $\lambda_b$ | Base failure rate (FIT). |
| $\pi_T$ | Temperature factor. |
| $\pi_A$ | Application factor. |

| Symbol | Description |
|---|---|
| $\pi_Q$ | Quality factor. |
| $\pi_E$ | Environment factor. |
| $\pi_S$ | Electrical stress factor. |
| $\pi_C$ | Contact construction factor. |
| $\pi_{CP}$ | Capacitance factor. |
| $\pi_V$ | Voltage stress factor. |
| $\pi_{SR}$ | Series resistance factor. |
| $\lambda_{NPC}$ | Failure rate of inverter (FIT). |
| $L$ | Lifetime under use condition. |
| $L_0$ | Lifetime under testing condition. |
| $V$ | Voltage under use condition (V). |
| $V_0$ | Voltage under testing condition (V). |
| $E_a$ | Activation energy (J). |
| $K_B$ | Boltzmann's constant ($8.62 \times 10^{-5}$ eV/K). |
| a | Constant describing the voltage and temperature dependency of $E_a$. |
| $\xi$ | Stress variable under operation. |
| $\xi_0$ | Stress variable under test. |
| $P_r$ | Rated power of MOSFETs (W). |
| $V_S$ | Voltage Stress Ratio. |
| $T_j$ | Junction temperature (°C). |
| $T_a$ | Ambient temperature (°C). |
| $T_C$ | Case temperature (°C). |
| $T_H$ | Heat sink temperature (°C). |
| $R_{th,ca}$ | Thermal resistance between the case and ambient (°C/W). |
| $R_{jC}$ | Thermal resistance between the junction and case (°C/W). |
| $R_{th,cH}$ | Thermal resistance between the case and heat sink (°C/W). |
| $R_{th,Ha}$ | Thermal resistance between the heat sink and ambient (°C/W). |
| $Z_{jC}$ | Thermal impedances between the junction and case (°C/W). |
| $P_{SW,MOSFET}$ | Switching losses of MOSFET (W). |
| $P_{SW,D}$ | Switching losses of diode (W). |
| $T_S$ | Sampling period (Sec). |
| $E_{on}$ | Turn on energy losses (J). |
| $E_{off}$ | Turn off energy losses (J). |
| $E_{REC}$ | Reverse recovery process energy (J). |
| $E(M,\theta)$ | Commutation energy losses (J). |
| $I_l(M,\theta)$ | Load current (A). |
| $I_{max}$ | Maximum collector current (A). |
| $M$ | Modulation index. |
| $\varphi$ | Current lagging angle to voltage (Deg). |
| $V_{CE}$ | Collector to emitter voltage (V). |
| $V_{CEN}$ | Rated collector to emitter voltage (V). |
| $V_{CEO}$ | Threshold collector to emitter voltage (V). |
| $I_C$ | Collector current (A). |
| $I_{CN}$ | Rated collector current (A). |
| $R_S$ | Collector to emitter resistance (Ω). |
| $V_F$ | Diode forward voltage (V). |
| $V_{FN}$ | Rated diode forward voltage (V). |
| $V_{FO}$ | Diode threshold voltage (J). |
| $R_D$ | Diode resistance (Ω). |
| $P_{cond}$ | Conduction losses (W). |
| $E_{cond}$ | Conduction energy (J). |
| α | Command voltage vector angle (Deg). |
| $\lambda_{D,F}$ | Failure rate of freewheeling diode (FIT). |
| $\lambda_{D,C}$ | Failure rate of clamping diode (FIT). |

## 1. Introduction

Nowadays, continuous development of semiconductor switches has led to widespread application of power electronic systems. Usually, these systems have a large number of power semiconductor switches. In addition, most of power electronic converters are equipped with electrolytic capacitors. But the semiconductor switches and electrolytic capacitors are the most fragile components [1]-[3]. Also, cost reduction pressure from global competition dictates minimum reliability-oriented design margin. For these reasons, reliability is the number one challenge for power electronic systems. So, quantitative evaluation of reliability for power electronic systems being a significant concern, can be used as a criterion to compare different topologies and control strategies.

The past decade has witnessed an increasingly growing research interest in various aspects of reliability for power electronic systems, with focused specifically on inverters [4]–[7]. In [8], Chiodo *et al.* presented some crucial properties to evaluate reliability of the power electronic systems. During the last few decades, many recommendations are proposed to improve reliability, such as "fault-tolerant design", "condition monitoring", and "active thermal management", to meet current and future industry needs. Mirafzal presented an instructive survey of existing fault-tolerance techniques for three-phase, two-level, and multilevel inverters in [9]. More comprehensive fault-tolerant techniques regarding power electronic converters in case of power semiconductor device failures, are reviewed by [10]. For condition monitoring (CM), a review paper was presented by [11], which described the current state of the art in CM research for power electronics. In [12], it is proposed to use the active thermal management to reduce the switching losses or to move them to less stressed devices. That can increase the reliability of power electronic modules. In [13], the authors present a global reliability comparison between two-level and three-level/five-level inverter topologies in single and three-phase operations. Harb *et al.* has proposed a new methodology for calculating the reliability of the photovoltaic module-integrated inverter (PV-MII) based on a stress factor approach [14]. Various fault-tolerant configurations have been proposed in the literature for power electronics converters [15]–[20]. But, no reliability evaluation or comparisons of different control strategies have been presented in previous articles. For the first time to our knowledge, a general methodology is applied that permits us to compare different control strategies from the reliability point of view. Though the methodology presented here is general, results associated with a three-phase three-level neutral point clamped (NPC) inverter are presented and discussed here.

NPC inverters are the most widely used topology of multilevel inverters in MV high-power applications on the market [21] and play an increasingly important role in electric motor speed control, utility interfaces with renewable energy resources, induction heating, flexible AC transmission systems (FACTS) and uninterruptible power supplies (UPS). Over the last decades, many modulation schemes are proposed to improve the performance of NPC inverters, which can generally be classified into two categories: pulse width modulation (PWM), and space vector modulation. In this paper, three common control methods: sinusoidal PWM (SPWM), third harmonic injection PWM (THIPWM), and space vector PWM (SVPWM) are compared to determine the most reliable modulation technique for NPC inverter. It is demonstrated that the effect of control strategy on reliability will be determined by its effect on the junction temperature of switches. The junction temperature is a function of three parameters, namely, ambient temperature, power loss of the semiconductor switch and the thermal resistance of the heat sink. Clearly, control strategy effects on the junction temperature only through the power loss. So this paper presents a comprehensive analysis on conduction and switching loss for different modulation strategies.

## 2. Basic operation of the three-level NPC inverter

In NPC topology, to produce n different levels for output phase voltage, (n-1) capacitors (with DC voltages), 2(n-1) switches and (n-1)(n-2) clamping diodes are needed in each leg. Fig. 1 depicts the three-phase three-level NPC inverter topology. Switches (1, 3) and (2, 4) on each leg are a complementary switching pair, which means that when a switch is on, to avoid DC link short circuit, the other switch must be off and vice versa. Table 1 shows the three switching states of this topology and corresponding output voltage levels. Also, their corresponding equivalent circuit is highlighted in Fig. 2. To obtain the equation of power losses we need to calculate the effective duty cycle of each switching state. Table 2 shows the duty cycle of each state based on modulating function (MF). The modulating function is explained for different control strategies as a function of modulation index and current to voltage lagging angle, in Section 6.

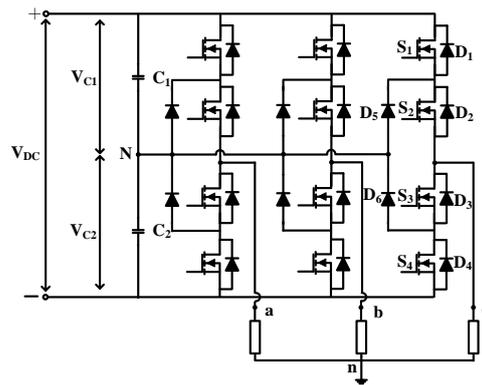

*Fig. 1.* *Three-phase three-level topology of a diode clamped inverter.*

**Table 1** The switching state of NPC inverter and corresponding output voltage levels

| State | $S_1$ | $S_2$ | $S_3$ | $S_4$ | NPC output voltage |
|-------|-------|-------|-------|-------|--------------------|
| 1 | ON | ON | OFF | OFF | $V_{DC}/2$ |
| 2 | OFF | ON | ON | OFF | 0 |
| 3 | OFF | OFF | ON | ON | $-V_{DC}/2$ |

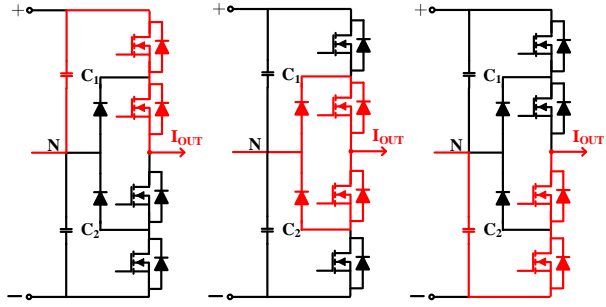

*Fig. 2. Corresponding equivalent circuit for switching states.*

**Table 2** The Duty Cycle of Switching States

| Switching State | Effective Duty Cycle | Symbol |
|---|---|---|
| 1 Positive Output | MF | $DT_P$ |
| 2 Zero Output in First Half Cycle | 1-MF | $DT_{ZP}$ |
| 2 Zero Output in Second Half Cycle | 1+MF | $DT_{ZN}$ |
| 3 Negative Output | -MF | $DT_N$ |

## 3. Failure rate of components

In order to analysis the effect of modulation schemes on the reliability of the components, this section is devoted to the calculation of the failure rate of the components that is relevant to this study. Currently, failure rates provided by the *Military Handbook for Reliability Prediction of Electronic Equipment*, MIL-HDBK-217 F [22], are used most often for the reliability modeling [23]. It covers the broad range of component types, and widely accepted for military and commercial electronic systems. So, in this paper MIL-HDBK-217 F will be used for failure rate calculations. However, the aim of this paper is to prepare a framework for reliability comparison of control strategies, and any available data source can be adopted in the outlined procedure. The field experience confirms that power switches and capacitors are the most vulnerable components. Moreover, magnetic components and control system are much more reliable [1], [13], [24], and [25]. Therefore, only power switches and capacitors are considered in this paper and other electronic systems (e.g. gate drivers, control) are not taken into account. In addition, the type of the components must be specified in this regard. In this paper the switches are considered to be power MOSFETs. But, a similar discussion can be extended to other types of power semiconductor switches. The failure rates of the MOSFET, diode, and capacitor are summarized in Table 3. These failure rates are expressed as a function of various stress factors.

**Table 3** Failure Rate of Components

| Part type | Failure rate ($10^{-9}$ failure/hour) | Stress factors |
|---|---|---|
| Capacitor | $\lambda_C = \lambda_b \pi_T \pi_{CP} \pi_V \pi_{SR} \pi_Q \pi_E$ | $\lambda_b = 0.00012$<br>$\pi_T = \exp\left(-4062\left(\frac{1}{T_a+273} - \frac{1}{298}\right)\right)$<br>$\pi_{CP} = C^{0.23}$<br>$\begin{cases} \pi_V = \left(\frac{S}{0.6}\right)^5 + 1 \\ S = \frac{V_{DC,Applied} + \sqrt{2}V_{AC,Applied}}{0.6 V_{Rated}} \end{cases}$<br>$\pi_{SR} = 1$<br>$\pi_Q = 10$<br>$\pi_E = 1$ |
| Diode | $\lambda_D = \lambda_b \pi_T \pi_S \pi_C \pi_Q \pi_E$ | $\lambda_b = 0.025$<br>$\pi_T = \exp\left(-3091\left(\frac{1}{T_j+273} - \frac{1}{298}\right)\right)$<br>$\begin{cases} \pi_S = 0.054 & \text{For } V_S \leq 0.3 \\ \pi_S = V_S^{2.43} & \text{For } 0.3 \leq V_S \\ V_S = \frac{V_{rev,applied}}{V_{rev,rated}} \end{cases}$<br>$\pi_C = 1$<br>$\pi_Q = 8$<br>$\pi_E = 1$ |
| MOSFET | $\lambda_{FET} = \lambda_b \pi_T \pi_A \pi_Q \pi_E$ | $\lambda_b = 0.012$<br>$\pi_T = \exp\left(-1925\left(\frac{1}{T_j+273} - \frac{1}{298}\right)\right)$<br>$\pi_A = 8 \quad \text{For } 50^W \leq P_r \leq 250^W$<br>$\pi_Q = 8$<br>$\pi_E = 1$ |

As shown in Table 3, except the voltage stress factor, other factors are equal in each control strategy for the failure rate of capacitors. Also, the junction temperature is used as a common input for the failure rate calculation of the power switch. So, in the next three Sections, we will discuss about these factors in detail.

## 4. Failure rate of DC-link capacitors

DC link capacitors are widely used to balance the instantaneous power difference between the input source and output load, and minimize voltage variation in the dc link. Three types of capacitors are generally available for dc-link applications, which are the aluminum electrolytic capacitors, metalized polypropylene film capacitors and high capacitance multi-layer ceramic capacitors [26]-[28]. Among these types, aluminum electrolytic capacitors due to low cost per joule are commonly used in DC-link application. So, in this paper dry electrolytic aluminum type is considered for capacitors. This choice is purely for illustrative purposes. Failure-rate models for other technologies are available in [22] and can be incorporated into the analysis for comparing different technologies.

The failure rate of capacitor can be calculated using many lifetime models. Arrhenius equation-based models are most widely employed to analyze the reliability of capacitors. These models are generally expressed as follows:

$$L = L_0 \times \left(\frac{V}{V_0}\right)^{-n} \times e^{\left(\frac{E_a}{K_B}\right)\left(\frac{1}{T} - \frac{1}{T_0}\right)} \quad (1)$$

The $E_a$ and $n$ as a function of the capacitor type were obtained in [29], [30]. For aluminum electrolytic capacitors, equation (1) can be simplified as follows [31]:

$$L = L_0 \times \left(\frac{V}{V_0}\right)^{-n} \times 2^{\frac{T_0 - T}{10}} \quad (2)$$

In [32], a generic lifetime model of electrolytic capacitors is proposed based on the primary wear-out mechanism of electrolyte as follows:

$$L = \begin{cases} L_0 \times \dfrac{V}{V_0} \times e^{\left(\frac{E_a}{K_B}\right)\left(\frac{1}{T} - \frac{1}{T_0}\right)} & \text{(low } \xi\text{)} \\ L_0 \times \left(\dfrac{V}{V_0}\right)^{-n} \times e^{\left(\frac{E_a}{K_B}\right)\left(\frac{1}{T} - \frac{1}{T_0}\right)} & \text{(medium } \xi\text{)} \quad (3) \\ e^{a_1(V_0 - V)} \times e^{\left(\frac{E_{a0} - \xi a_0}{K_B T} - \frac{E_{a0} - \xi a_0}{K_B T_0}\right)} & \text{(high } \xi\text{)} \end{cases}$$

In this paper, to integrate analytical tools for assessing converter reliability, the MIL-HDBK-217 F will be used to calculate the failure rate of capacitors.

As previously mentioned, except the voltage stress factor, other factors are equal in each control strategy for failure rate of capacitors. To calculate the voltage stress, the voltage of DC link capacitors for different modulation techniques are obtained (see Fig. 3). Based on these results, it is clear that the failure rates of $C_1$, $C_2$ in SVPWM modulation are different. This is because of imbalance voltages of DC-link capacitors, which causes the sum of the failure rates of $C_1$ and $C_2$ to be higher in this method compared to the other two methods. Although some improved SVPWM methods are proposed to reduce the voltage imbalance, but these methods have a stronger impact on imbalance losses in power switches [33]. So, usually these methods decrease the reliability. It should be noted that beside the voltage balancing, voltage ripple reduction has an equal importance from the power electronic designers perspective and more research efforts are expected to tackle these issues to achieve more reliable inverters.

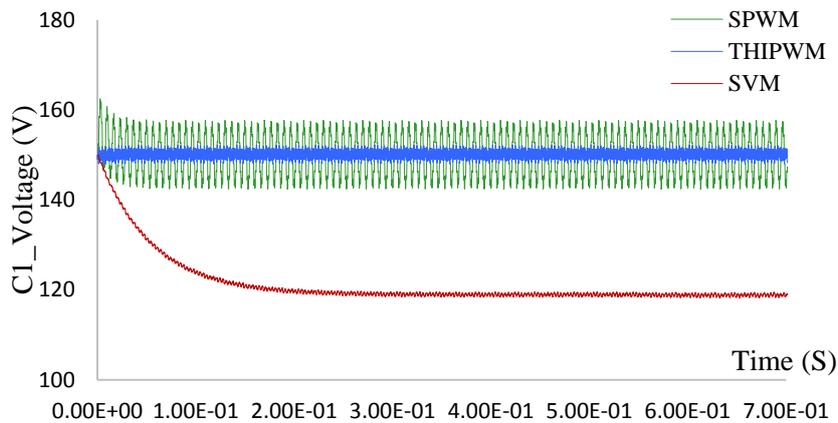

a

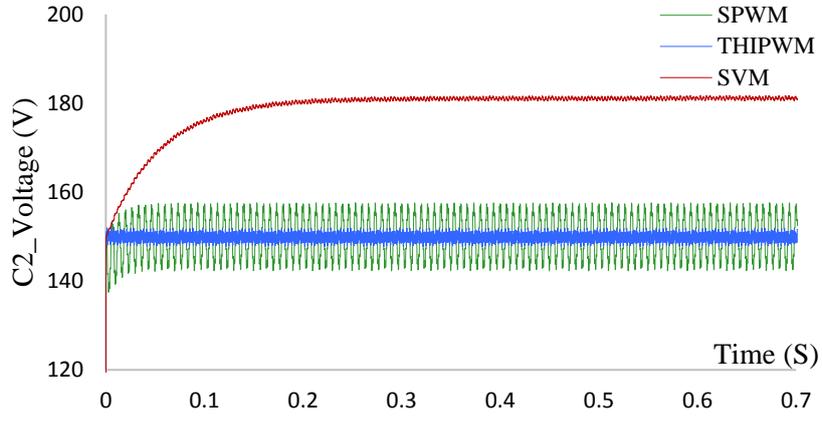

b

*Fig. 3.* *Capacitors voltage comparison with different control strategies*
a $C_1$ voltage
b $C_2$ voltage

## 5. Junction temperature calculation

As it was mentioned before, the temperature of the switch is the only factor that affects the failure rate of the switches in different control strategies. Consequently, in this section the thermal modeling of the switches is provided. The thermal models used for a single power switch and a power switch module are shown in Fig. 4, in which the thermal impedance between the junction and case usually is modeled as a multi-layers foster RC network in the manufacturer datasheets, (see Fig. 5c) [34], [35]. Regardless of the thermal capacitance $C_{th}$, which describe dynamic changes, junction temperature based on the thermal equivalent model shown in Fig. 4, is calculated as follows:

$$T_j = T_C + P_{loss} R_{th,jC} \quad (4)$$

Similar to (4), the case temperature of switch can be expressed in terms of its power loss and ambient temperature as

$$T_C = T_a + P_{loss} R_{th,Ca} \quad (5)$$

Using (4) and (5), the junction temperature can be expressed as

$$T_j = T_a + P_{loss}(R_{th,jC} + R_{th,Ca}) \quad (6)$$

So having the ambient temperature, power losses and thermal resistance we can calculate the junction temperature of power switches. Note that, if a heat sink is used, the thermal resistance between the case and ambient will be the thermal resistance of the heat sink, which is much less than the thermal resistance between the case and ambient when no heat sink is used [36].

It should be noted that junction temperature is not very sensitive to ambient temperature. Also, for a specific application with known power loss, the heat sink will be predetermined. Therefore, power loss change is the only factor that leads to a temperature change. The next section is devoted to the power loss calculation for different control strategies.

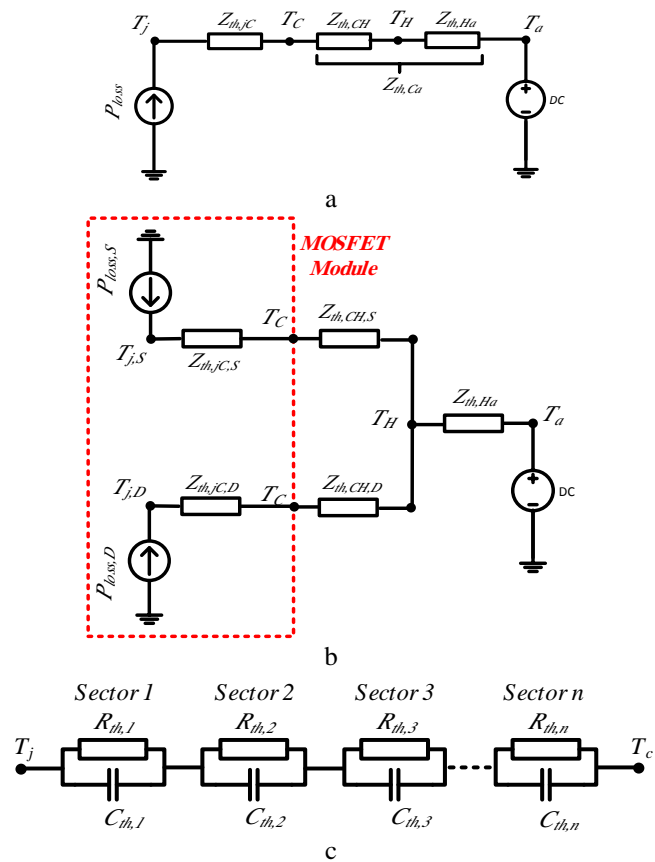

***Fig. 4.*** *The used thermal model and thermal equivalent for $Z_{th,jC}$*
a The used thermal model for single power switch
b The used thermal model for power switch with freewheeling diode module
c Thermal equivalent for $Z_{th,jC}$ (Foster network)

# 6. Evaluation of power switches losses

Power switch losses consist of conduction losses and switching losses. Up to now, many papers have reported conduction and switching loss calculation, but data sheet information based method for switching losses calculation is well known and widely accepted in both scientific and industrial applications. In this method, characteristic curves which are presented in the datasheets of each power semiconductor, are approximated by exponential equations using curve-fitting tools. This method provides an accurate power loss prediction for many different types of circuits. On the other hand, power loss prediction based on pulse by pulse calculation is used most often for the conduction loss calculation. It can be seen that the losses are dependent on the circuit parameters. In addition to this, it is easy to use in order to compare conduction loss of control techniques.

## 6.1. Switching Losses

Switching loss consists of the energy losses during turn-on and turn-off instants in one reference period. Turn-on losses are caused by the forward recovery process. As for fast diodes, this share of the losses can be neglected. However, the switching energy at turn-off can't be neglected. The switching losses for power switches and diodes can be derived as:

$$P_{SW,MOSFET} = f_S \left( E_{ON} \left( I_l(M,\theta) \right) + E_{OFF} \left( I_l(M,\theta) \right) \right) \quad (7)$$

$$P_{SW,D} = f_S \times E_{REC} \left( I_l(M,\theta) \right) \quad (8)$$

Where the commutation energy loss as a function of load current is described as follows:

$$E(M,\theta) = A e^{B I_l(M,\theta)} + C e^{D I_l(M,\theta)} \quad (9)$$

Where (A, B) and (C, D) are turn-on and turn-off curve fitting constants for power switch, respectively. For a typical MOSFET (IRF740), the turn-on energy losses, turn-off energy losses, and reverse recovery process energy were computed with a more precise approximation with the following relations:

$$E_{ON} = 0.0048e^{0.0044 I_l(M,\theta)} - 0.00433e^{-0.008 I_l(M,\theta)} \quad (10)$$

$$E_{OFF} = 0.0126e^{-0.00107 I_l(M,\theta)} - 0.0102e^{0.00021 I_l(M,\theta)} \quad (11)$$

$$E_{REC} = 0.00806e^{-0.000322 I_l(M,\theta)} - 0.0057e^{-0.00446 I_l(M,\theta)} \quad (12)$$

$$I_l(M,\theta) = M\, I_{max} \sin(\theta - \phi) \quad (13)$$

In this section, to avoid losing any generality, load current is described as (13). However in practice, the load current data are extracted from the corresponding waveform then switching energy loss is calculated point by point via a MATLAB script program along time.

### 6.2. The conduction losses for SPWM and THIPWM

Conduction loss occurs during the on-state mode of power switch. The curve of collector to emitter voltage versus collector current is usually approximated by the following linear equation (this implies a threshold voltage plus a resistive voltage drop) [37]

$$V_{CE} = \frac{V_{CEN} - V_{CEO}}{I_{CN}} I_C + V_{CEO} = R_S I_C + V_{CEO} \quad (14)$$

Similarly, diode forward voltage versus diode current is described diagram as follows:

$$V_F = \frac{V_{FN} - V_{FO}}{I_{CN}} I_C + V_{FO} = R_D I_C + V_{FO} \quad (15)$$

During the power semiconductor operation, the conduction energy loss based on effective duty cycle in $S_1$ is calculated as follows:

$$E_{cond,S1} = (R_S\, I_{max} \sin(\omega t) + V_{CEO}).I_{max} \sin(\omega t).DT_P \quad (16)$$

Then the conduction power loss is defined as:

$$P_{cond,S1} = \frac{1}{T}\int dE_{cond,S1} = \frac{1}{T}\int_0^{\pi-\varphi} (R_S\, i_C + V_{CEO}).I_{max}\sin(\omega t).MF\, d\omega t \quad (17)$$

Also, the modulation functions for SPWM and THIPWM can be calculated according to the following equations [37], [38]:

$$MF_{SPWM} = M \cdot \sin(\omega t + \varphi) \quad (18)$$

$$MF_{THIPWM} = \frac{2M}{\sqrt{3}}\left[\sin(\omega t + \varphi) + \frac{1}{6}\sin(3\omega t + 3\varphi)\right] \quad (19)$$

By considering the above equations, conduction losses in SPWM and THIPWM schemes can be deduced as follows:

$$P_{cond,SPWM,S1} = \frac{MR_S I_{max}^2}{6\pi}(1+\cos\varphi)^2 +$$

$$\frac{M I_{max} V_{CEO}}{4\pi}((\pi-\varphi)\cos\varphi + \sin\varphi) \quad (20)$$

$$P_{cond,THI,S1} = \frac{M I_{max}}{180\sqrt{3}\pi}\left\{8 I_{max} R_S \cos^4\left(\frac{\varphi}{2}\right)(37-8\cos\varphi)\right.$$

$$\left. + 15 V_{CEO}\left[6(\pi-\varphi)\cos\varphi + (6+\sin^2\varphi)\sin\varphi\right]\right\} \quad (21)$$

The conduction losses for other switches and diodes are given in the Appendix.

### 6.3. The conduction losses expressions for SVPWM

In space vector pulse width modulation (SVPWM), expressing the waveforms is difficult. To solve this problem, some attempts have been done to find out the relations between carrier based PWM and SVPWM schemes. The most important of these methods include 1) injecting a common mode voltage with suitable magnitude to the sinusoidal reference phase voltage in the case of carrier based PWM, and 2) employing dwell times in equivalent redundant switching states in the case of space-vector modulation [36]. The voltage expression for various values of modulation index is different. Regions of space vector diagram which are based on different modulation indexes have unique expression as shown in Fig. 5, and their expressions are given in Tables 4-6. So similar to SPWM, we can calculate the conduction losses for SVPWM same as [39], [40].

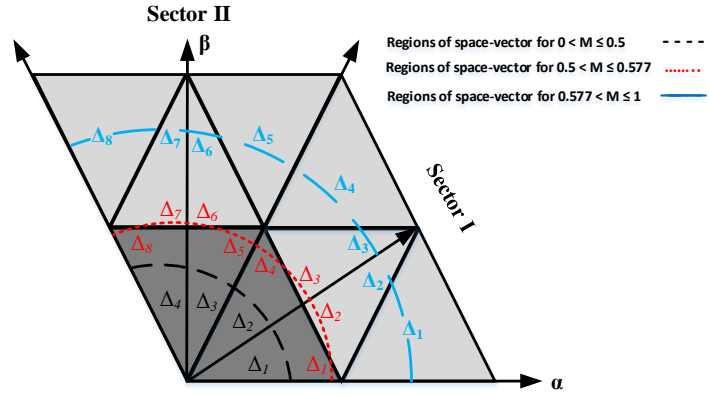

***Fig. 5.*** *Space vector diagram regions during 120 excursion of the reference for various values of modulation index.*

**Table 4** The Voltage Expression of SVPWM for $0 < M \leq 0.5$

| Region | Voltage expression for phase A | Region | Voltage expression for phase A |
|---|---|---|---|
| $\Delta_1$ | $\frac{M}{\sqrt{3}}\cos(\alpha+\frac{\pi}{6})$ | $\Delta_3$ | $\frac{M}{\sqrt{3}}\cos(\alpha-\frac{\pi}{6})$ |
| $\Delta_2$ | $M\cos\alpha$ | $\Delta_4$ | $\frac{M}{\sqrt{3}}\cos(\alpha+\frac{\pi}{6})$ |

**Table 5** The Voltage Expression of SVPWM for $0.5 < M \leq 0.577$

| Region | Voltage expression for phase A | Region | Voltage expression for phase A |
|---|---|---|---|
| $\Delta_1$ | $\frac{M}{\sqrt{3}}\cos(\alpha+\frac{\pi}{6})$ | $\Delta_5$ | $\frac{M}{\sqrt{3}}\cos(\alpha-\frac{\pi}{6})$ |
| $\Delta_2$ | $M\cos\alpha-0.25$ | $\Delta_6$ | $\frac{M}{\sqrt{3}}\cos(\alpha+\frac{\pi}{6})+0.25$ |
| $\Delta_3$ | $\frac{M}{\sqrt{3}}\cos(\alpha+\frac{\pi}{6})+0.25$ | $\Delta_7$ | $\frac{M}{\sqrt{3}}\cos(\alpha-\frac{\pi}{6})-0.25$ |
| $\Delta_4$ | $M\cos\alpha$ | $\Delta_8$ | $\frac{M}{\sqrt{3}}\cos(\alpha+\frac{\pi}{6})$ |

**Table 6** The Voltage Expression of SVPWM for $0.577 < M \leq 1$

| Region | Voltage expression for phase A | Region | Voltage expression for phase A |
|---|---|---|---|
| $\Delta_1$ | $\frac{M}{\sqrt{3}}\cos(\alpha-\frac{\pi}{6})$ | $\Delta_5$ | $M\cos\alpha$ |
| $\Delta_2$ | $M\cos\alpha-0.25$ | $\Delta_6$ | $\frac{M}{\sqrt{3}}\cos(\alpha+\frac{\pi}{6})+0.25$ |
| $\Delta_3$ | $\frac{M}{\sqrt{3}}\cos(\alpha+\frac{\pi}{6})+0.25$ | $\Delta_7$ | $\frac{M}{\sqrt{3}}\cos(\alpha-\frac{\pi}{6})-0.25$ |
| $\Delta_4$ | $\frac{M}{\sqrt{3}}\cos(\alpha-\frac{\pi}{6})$ | $\Delta_8$ | $M\cos\alpha$ |

## 7. Power loss comparison

Power losses for one of the switches ($S_1$), for different modulation strategies as a function of modulation index and lagging angle of current to voltage are shown in Fig. 6. As it can be seen in Fig. 6, THIPWM and SVPWM power losses of $S_1$ are approximately equal. Also, it is clear that the losses with SVPWM modulation is lower than other schemes. Consequently, the right side of the temperature factor equation decreases, i.e., junction temperature decreases as well. Which in turn leads to increase the life time of $S_1$. Thus, the SVPWM modulation is more appropriate for the $S_1$.

Distribution losses of $S_1$ under unit modulation index and unit power factor in line cycle is compared in Fig. 7a for different modulation strategies. It can be seen that conduction loss take a larger part of the whole losses. So, most efforts to reduce losses should be done in this part. It is noted that this result largely depends on the static characteristics of MOSFET. Distribution losses of NPC inverter are also compared in Fig. 7b. Switches have a considerable share of losses due to their characteristic. It can be seen that intermediate switches have more losses and they are more prone to failure.

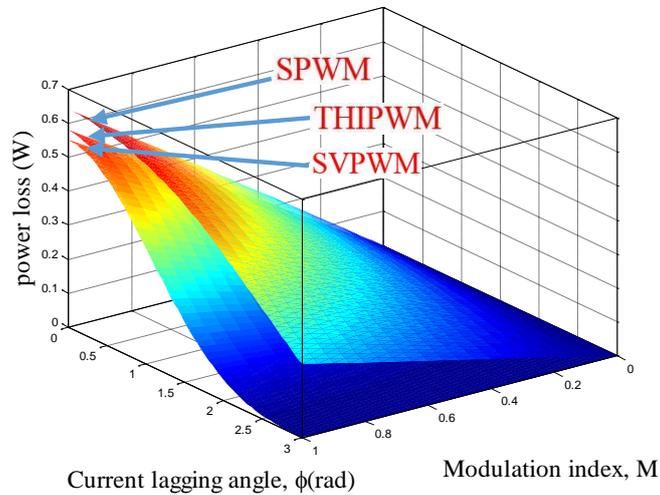

*Fig. 6. Power losses of $S_1$ for different modulation strategies as a function of modulation index and current lagging angle to voltage.*

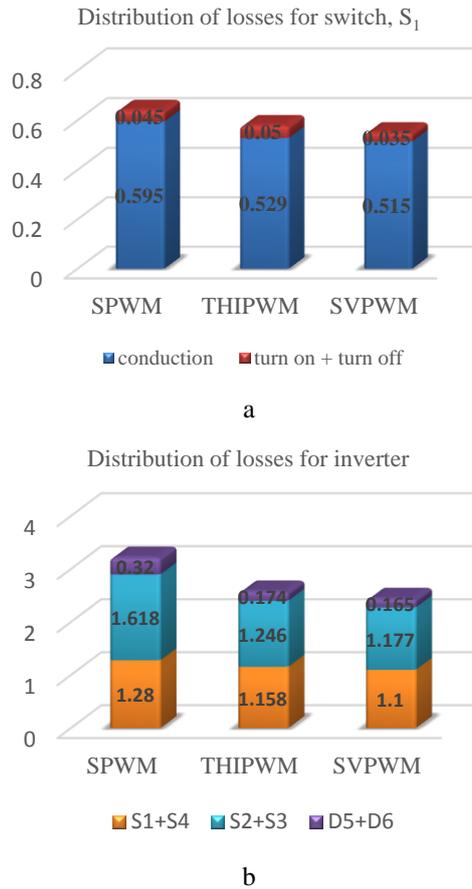

*Fig. 7. Distribution losses*
a Distribution losses of NPC inverter for different modulation strategies (M = 1, cos (φ) = 1)
b Distribution losses of $S_1$ for different modulation strategies (M = 1, cos (φ) = 1)

## 8. Experimental results

A three-phase three-level NPC inverter have been built for verifying the theoretical analysis presented above. The experimental setup is shown in Fig. 8. To make it easy to compare the simulation and experimental results, the parameters of experimental setup are the same as those of the simulation, which can be found in Table 9 in the Appendix. In order to increase the temperature faster, twelve individual power modules without heat-sink are used.

Fig. 9 shows the thermal image of the inverter with NPC topology at different modulation strategies, which instead of the inaccessible junction temperatures, represent case temperatures of the individual switches. Since the temperatures of the legs are so close to each other, only the left leg is shown in Fig. 9. As it is anticipated, Fig. 9 demonstrates a significant unbalance junction temperature distribution among the semiconductors. Also, it can be seen the junction temperature on the inner switches $S_2$ and $S_3$ is higher than on the outer switches, especially at SPWM modulation (Fig. 9-a). The case temperatures of switches and corresponding junction temperatures in each modulation strategy for left leg of NPC inverter are obtained and given in Table 7. A comparison between Table 7 and the junction temperatures obtained from simulations (by using (3) and Fig. 7b), shows that the experimental results are coincident with the simulation results.

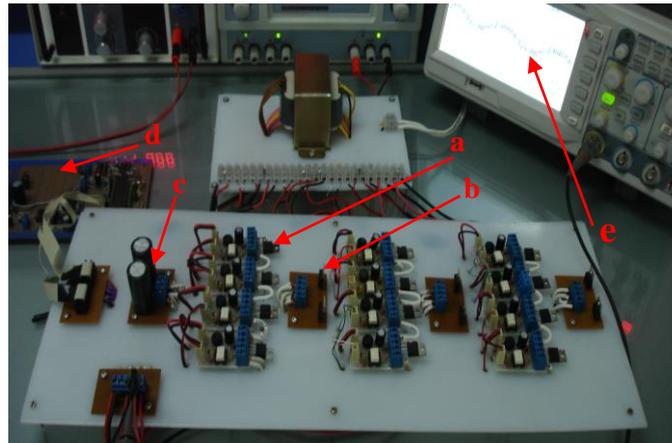

***Fig. 8.*** *Laboratory prototype of the three-phase three-level neutral point clamped (NPC) inverter. (a) Power switches (MOSFETs). (b) Clamping diodes. (c) DC link capacitors. (d) Controller (e) Line–line voltage.*

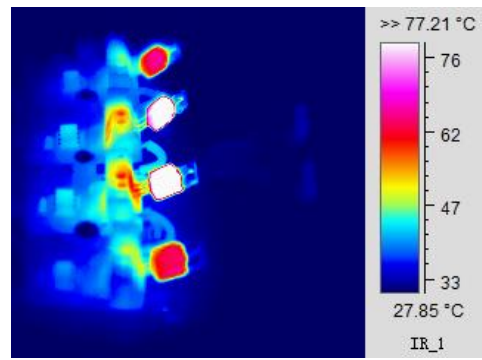

a

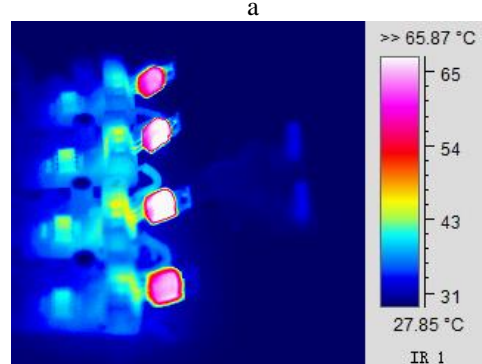

b

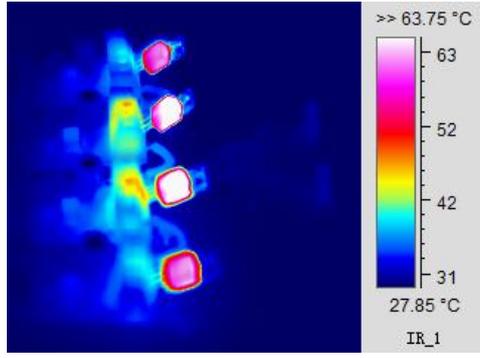
c

*Fig. 9. Thermal image of the NPC (Left leg)*
a With SPWM modulation
b With THIPWM modulation
c With SVPWM modulation

**Table 7** The Case Temperatures of Switches and Corresponding Junction Temperatures in Each Modulation Strategy for Left Leg of NPC Inverter

| Modulation /Component | | $S_1$ | $S_2$ | $D_5$ |
|---|---|---|---|---|
| SPWM | Case temperature | 64.04°C | 77.21°C | 34.52°C |
| | Junction temperature | 64.68°C | 78.06°C | 34.85°C |
| THIPWM | Case temperature | 60.37°C | 65.87°C | 30.17°C |
| | Junction temperature | 60.95°C | 66.54°C | 30.35°C |
| SVPWM | Case temperature | 58.57°C | 63.75°C | 29.91°C |
| | Junction temperature | 59.12°C | 64.38°C | 30.08°C |

## 9. Comparative reliability evaluation of control strategies

In this section, quantitative analysis will be done to demonstrate the effect of different control strategies on the lifetime of the inverter. Since, in a NPC inverter without fault tolerant capability, first failure occurrence will lead to fail the inverter. Thus, first failure is the critical failure; and system is operating if all components are working properly. So, the reliability model of this inverter is in series from the reliability point of view. Therefore the failure rate of the inverter can be expressed as:

$$\lambda_{NPC} = \sum_{i=1}^{12} \lambda_{\text{FET},i} + \sum_{j=1}^{12} \lambda_{D,F,j} + \sum_{k=1}^{6} \lambda_{D,C,k} + \sum_{L=1}^{2} \lambda_{C,L} \quad (22)$$

Failure rates in (22) and respective stress factors are calculated according to the previous sections and summarized in Table 8. These results are related to the mentioned three control strategies. With these data, it is possible to calculate the reliability of the NPC inverter for each modulation. As shown in Table 8, the most contribution stress factor to failure rates of power switches are related to temperature. So, the junction temperature of the switches is a key feature in determining the more reliable modulation technique. Fig. 10 shows the percentage contribution of each component in the inverter failure rate. Clearly, the most failure prone devices are the power switches. So, for reliability improvement, thermal management for switches based on junction temperature control is recommended. To this end, two solutions are proposed: 1) Using the Active NPC topology with a proper modulation strategy for better loss distribution in the semiconductor devices, in which distribution of losses are controlled by the selection of the different NPC paths at the zero states; 2) Active cooling methods could be applied to certain types of switches with high losses to reduce the junction temperature and therefore extend the lifetime of the switches.

**Table 8** Stress Factors and Predicted Failure Rate for Each Component

|  |  | MOSFETs and anti-parallel diodes | | Diodes clamped | Capacitors | |
|---|---|---|---|---|---|---|
|  |  | $S_1, S_4$ | $S_2, S_3$ | $D_1, D_2, D_3, D_4$ | $D_5, D_6$ | $C_1$ | $C_2$ |
| $\lambda_b$ ($10^{-6}h^{-1}$) | | 0.012 | 0.012 | 0.025 | 0.025 | $12\times10^{-5}$ | $12\times10^{-5}$ |
| $\pi_A$ | | 8 | 8 | - | - | - | - |
| $\pi_S$ | | - | - | 0.19 | 0.19 | - | - |
| $\pi_C$ | | - | - | 1 | 1 | - | - |
| $\pi_Q$ | | 8 | 8 | 8 | 8 | 10 | 10 |
| $\pi_E$ | | 1 | 1 | 1 | 1 | 1 | 1 |
| $\pi_{CP}$ | | - | - | - | - | 4.12 | 4.12 |
| $\pi_{SR}$ | | - | - | - | - | 1 | 1 |
| $\pi_V$ | SPWM | - | - | - | - | 13.61 | 13.61 |
|  | THIPWM | - | - | - | - | 10.90 | 10.90 |
|  | SVPWM | - | - | - | - | 4.15 | 26.02 |
| $\pi_T$ | SPWM | 2.136 | 2.655 | 1.103 | 1.394 | 2.872 | 2.872 |
|  | THIPWM | 2.005 | 2.204 | 1.103 | 1.201 | 2.872 | 2.872 |
|  | SVPWM | 1.942 | 2.125 | 1.103 | 1.190 | 2.872 | 2.872 |
| Predicted failure rate ($10^{-6}h^{-1}$) | SPWM | 1.640 | 2.039 | 0.042 | 0.053 | 0.193 | 0.193 |
|  | THIPWM | 1.537 | 1.692 | 0.042 | 0.047 | 0.155 | 0.155 |
|  | SVPWM | 1.491 | 1.632 | 0.042 | 0.045 | 0.059 | 0.369 |

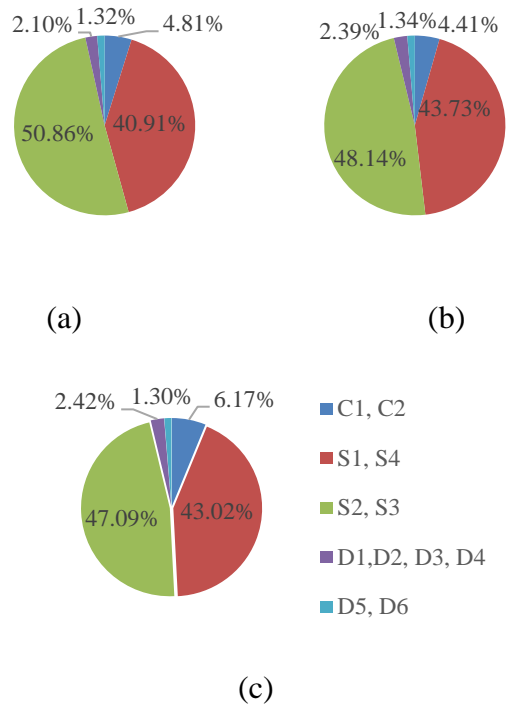

*Fig. 10. Percentage contributions from the each component to the inverter failure rate. (a) SPWM. (b) THIPWM. (c) SVPWM.*

For reliability comparison between modulation methods, the mean time to failure (MTTF) is applied, which gives the length of time a device or other product is expected to last in operation. That by the inverse of the failure rate function, MTTF can be calculated:

$$MTTF = \frac{1}{\lambda_{NPC}} \quad (23)$$

By taking the above equation and Table 8, MTTF of NPC inverter with SPWM, THIPWM and SVPWM are 0.042951 ($10^6$h), 0.048852 ($10^6$h) and 0.050135 ($10^6$h), respectively. It is clear that MTTF of inverter with SVPWM modulation stage is higher than other modulations (13.74 percent higher than THIPWM and 16.73 percent higher than SPWM). Therefore, SVPWM modulation not only provides excellent output performance but also improves the reliability level.

## 10. Conclusion

For the first time, a comprehensive reliability analysis between the SPWM, THIPWM and SVPWM modulations has been carried out. The lifetime of the NPC inverter is obtained and compared for each control strategy. It was shown that the junction temperature of the switches is a key feature in determining the more reliable modulation. To calculate the junction temperature, switch loss calculation in a NPC inverter was investigated for different modulations. This comparison shows that the SVPWM strategy has the highest reliability for the NPC topology.

The results confirm that power switches are the most assailable components. So, further research may investigate the control strategy with a loss distribution to increase reliability. In addition, the method presented in this paper can be extended to other inverter topologies.

The reader is undoubtedly aware that design choices such as the choice of capacitor technology, switch ratings, and power factor can influence the reliability of system. So, the goal of this paper was not to judge the reliability merits of one control strategy over another, but rather to present a systematic framework of applying a usage model for comparing different control strategies on the basis of MTTF.

## 12. Appendices

The conduction loss expressions for SPWM modulation:

$$P_{cond,S2} = P_{cond,S3} = \left\{ \frac{R_S I_{max}^2}{4} + \frac{I_{max} V_{CEO}}{\pi} \right\}$$

$$-\frac{M R_S I_{max}^2}{6\pi}(1-\cos\varphi)^2 + \frac{M I_{max} V_{CEO}}{4\pi}(\varphi\cos\varphi - \sin\varphi)$$

$$P_{cond,S4} = P_{cond,S1}$$

$$P_{cond,D1} = P_{cond,D2} = P_{cond,D3} = P_{cond,D4}$$

$$P_{cond,D1} = \frac{MV_{FO}\,I_{max}}{4\pi}(\sin\varphi - \varphi\cos\varphi) + \frac{MR_D I_{max}^2}{6\pi}(1-\cos\varphi)^2$$

$$P_{cond,D5} = P_{cond,D6} = \left\{ \frac{R_D I_{max}^2}{4} + \frac{I_{max} V_{FO}}{\pi} \right\}$$

$$+ M \left\{ \frac{I_{max} V_{FO}}{4\pi}[(2\varphi-\pi)\cos\varphi - 2\sin\varphi] - \frac{R_D I_{max}^2}{3\pi}(1+\cos^2\varphi) \right\}$$

The conduction loss expressions for THIPWM modulation:

$$P_{cond,S2} = \frac{I_{max}}{540\pi}\left\{ -269\sqrt{3}\,MI_{max} R_S \sin^4\left(\frac{\varphi}{2}\right) \right.$$

$$+2\sqrt{3}\,M\cos\varphi[45\varphi V_{CEO} - 32 I_{max} R_S \sin^4\left(\frac{\varphi}{2}\right) + 15(9\pi I_{max} R_S$$

$$\left. +36 V_{CEO} - 6\sqrt{3}MV_{CEO} \sin\varphi - \sqrt{3}MV_{CEO} \sin^3\varphi)] \right\}$$

$$P_{cond,D1} = \frac{M\,I_{max}}{180\sqrt{3}\pi}\left\{ 269 I_{max} R_D \sin^4\left(\frac{\varphi}{2}\right) + \cos\varphi[-90\varphi V_{FO} \right.$$

$$\left. +64 I_{max} R_D \sin^4\left(\frac{\varphi}{2}\right)] + 15V_{FO}(6+\sin^2\varphi)\sin\varphi \right\}$$

$$P_{cond,D5} = \frac{I_{max}}{1080\pi}\left\{ -180\sqrt{3}M(\pi-2\varphi)V_{FO}\cos\varphi \right.$$

$$-84\sqrt{3}MI_{max} R_D \cos(2\varphi) + 5[-76\sqrt{3}MI_{max} R_D + 54\pi I_{max} R_D$$

$$\left. +216V_{FO} - 81\sqrt{3}MV_{FO}\sin\varphi + 3\sqrt{3}MV_{FO}\sin(3\varphi)] \right\}$$

**Table 9** Circuit Parameters

| Quantity | Value |
|---|---|
| DC-Link voltage | $300^V$ |
| Output frequency | $50^{Hz}$ |
| Carrier frequency | $1^{kHz}$ |
| DC-Link capacitors | $470^{\mu F}$ |
| Power factor | 1.0 |
| Modulation index | 1.0 |
| Switches | IRF740 |
| Thermal resistance between the junction and case for switches | $1^{\circ C/W}$ |
| Thermal resistance between case and ambient (in free air) for | $61^{\circ C/W}$ |
| switches | |
| Clamping diodes | MUR1560 |
| Thermal resistance between the junction and case for diodes | 2°C/W |
| Thermal resistance between case and ambient (in free air) for diodes | 58°C/W |
| Simulation step | 1 µs |